\documentstyle[preprint,aps,epsfig]{revtex}

\newcommand{\lsim}{\mathrel{\mathop{\kern 0pt \rlap 
  {\raise.2ex\hbox{$<$}}} 
  \lower.9ex\hbox{\kern-.190em $\sim$}}} 
\newcommand{\gsim}{\mathrel{\mathop{\kern 0pt \rlap 
  {\raise.2ex\hbox{$>$}}} 
  \lower.9ex\hbox{\kern-.190em $\sim$}}}

\newcommand{\beq}{\begin{equation}}
\newcommand{\eeq}{\end{equation}}
\newcommand{\be}{\begin{eqnarray}}
\newcommand{\ee}{\end{eqnarray}}

\begin{document} 

\draft 
\preprint{
\begin{tabular}{l}
\hbox to\hsize{March, 2003\hfill KIAS-P02054}\\[-2mm]
\end{tabular}
}

\title{ 
CP Violation in $B \to \phi K$ Decay \\
with Anomalous Right-handed Top Quark Couplings 
} 

\author{ 
Jong-Phil Lee$^{a}$ 
and 
Kang Young Lee$^{b}$\thanks{kylee@kias.re.kr}
} 

\address{ 
$^a$Department of Physics and IPAP, Yonsei University, Seoul 120-749, Korea\\ 
$^b$School of Physics, 
Korea Institute for Advanced Study, Seoul 130-012, Korea\\ 
} 

\maketitle 

\begin{abstract} 
\noindent 
We explore the CP violation in $B \to \phi K$ decay processes
in the presence of the anomalous right-handed 
$\bar{t}sW$ and $\bar{t}bW$ couplings.
The complex anomalous top coupling can be a source
of the new CP violation and lead to a deviation of
the observed weak phase in $B \to \phi K$ decays, 
which takes account for the present disagreement of the observed
$\sin 2 \beta$ between $B \to J/\psi K$ and $B \to \phi K$ decays.
The direct CP violation is also predicted.

\end{abstract} 
 
 
\tightenlines
 
\section{Introduction} 

Recently BaBar \cite{babar1} and Belle \cite{belle1} collaborations
report the first measurement of the time-dependent CP asymmetry
in $B \to \phi K$ decay to measure the weak phase $\sin 2 \beta$:
\be
\sin 2\beta &=& -0.73 \pm 0.64 \pm 0.18 ~~~~~~({\rm Belle}),
\nonumber \\
\sin 2\beta &=& -0.19^{+0.52}_{-0.50} \pm 0.09 ~~~~~~~~~~({\rm BaBar}),
\ee
where $\beta \equiv {\rm arg}(-V_{cd}V_{cb}^*/V_{td}V_{tb}^*)$.
In the Standard Model (SM), the origin of CP violation 
is only the complex phase of the Cabibbo-Kobayashi-Maskawa (CKM) 
quark mixing matrix.
It implies that $\sin 2 \beta$ in $B \to \phi K$ decays  
should agree with that of $B \to J/\psi K$ decays
up to small pollution of ${\cal O}(\lambda^2) \sim $ 5\% \cite{grossman,soni}.
Therefore a sizable disagreement of $\sin 2 \beta$ between 
$B \to \phi K$ and $B \to J/\psi K$ decays 
is a clear indication of new physics beyond the SM.
The world average of $\sin 2 \beta$ measured in $B \to J/\psi K$ decays
is given by \cite{nir}
\be
\sin 2\beta = 0.734 \pm 0.054,
\ee
which is consistent with the SM prediction and
indicates the nonzero CP violation in the $B$ system.
Remarkably, however, the measured $\sin 2 \beta$ 
in $B \to \phi K_S$ channel is far from that
of $B \to J/\psi K$ decay and even the central value is negative 
as shown in Eq. (1). 
At present, we confront a 2.7$\sigma$ discrepancy 
between average values of $\sin 2 \beta$ in 
$B \to \phi K_S$ and in $B \to J/\psi K_S$ decays.
Although it is premature to regard this disagreement 
as an evidence of the new physics due to a large statistical error,
the difference is so large that it is tempting to interpret it 
as a clue of the new physics.
Studies in various models are being performed 
to account for the discrepancy \cite{nir,new}.

The left-right (LR) model based on SU(2)$_L \times$ SU(2)$_R \times$ U(1)
gauge group is one of the natural extensions of the SM \cite{LR}.
In the LR model, 
the right-handed quark mixing is also an observable
as well as the left-handed quark mixing.
Without manifest symmetry
between left- and right-handed sectors,
the right-handed quark mixing is not necessarily same
as the left-handed quark mixing described by the CKM matrix.
Thus we have additional right-handed charged current interactions
with couplings different from left ones,
which are suppressed by the heavy mass of extra $W$ boson.
The strength of right-handed couplings should be determined
by measurements in various phenomena.
On the other hand, when the electroweak symmetry is dynamically broken,
some nonuniversal interactions may exist 
which lead to additional right-handed and left-handed couplings 
on charged current interactions \cite{zhang}.
If the anomalous right-handed $\bar{t}bW$ couplings exist,  
their effects can be found in rare $B$ decays \cite{larios,elhady}
and also in the various phenomena at future colliders \cite{boos,rindani}.

The production of $10^7 - 10^8$ top quark pairs per year
expected at Large Hadron Collider (LHC)
will allow us to study the structure of top quark couplings.
The $\bar{t}bW$ coupling will be directly measured with high precision
through the dominant $t \to b W$ channel
and the anomalous $\bar{t}bW$ coupling will be tested in direct way.
The subdominant channel of the top quark is
the CKM nondiagonal decay $t \to s W$ in the SM
of which branching ratio is estimated as
Br$(t \to s W) \sim 1.6 \times 10^{-3}$,
when $|V_{ts}| = 0.04$ is assumed.
Despite the small branching ratio of this channel,
the large number of expected top quark production at LHC
will enable us to measure the $t \to s W$ process
and provide us a chance to probe the $\bar{t}sW$ coupling directly.
Hence the anomalous $\bar{t}sW$ coupling
is worth studying at present.
In this work, we consider 
the anomalous right-handed $\bar{t}bW$ and $\bar{t}sW$ couplings
and their impact on the CP violation in $B \to \phi K_S$ decay.

We do not specify the underlying model here but
concentrate on the anomalous right-handed couplings of 
charged current interactions,
ignoring effects of additional left-handed interactions 
and new particles.
The relevant right-handed couplings are described by 
the effective lagrangian as
\be
{\cal L} = -\frac{g}{\sqrt{2}}
          \sum_{q=s,b} V_{tq}~ \bar{t} \gamma^\mu
                                 (P_L + \xi_q P_R) q W^+_\mu
  + H.c.,
\ee
where $\xi_q$ is dimensionless parameter measuring new physics effects.
If $\xi_q$ has a complex phase, generically it invokes a new CP violation 
leading to the shift of the observed $\sin 2 \beta$.

This paper is organized as follows:
In section II, the effective Hamiltonian formalism with right-handed 
$\bar{t} b W$ and $\bar{t} s W$ couplings is given.
In section III, we discuss the constraints on the parameters $\xi_{b}$
and $\xi_{s}$ using the radiative $B \to X_s \gamma$ decay.
The analysis on hadronic decays $B \to J/\psi K$ and
$B \to \phi K$ is presented in section IV 
to extract corresponding $\sin 2 \beta$. 
Finally we conclude in section V.

\section{The Effective Hamiltonian}

The effective Hamiltonian approach is required
when we study rare decays of $B$ mesons in order to incorporate
QCD effects in a consistent way.
The $\Delta B =1$ effective Hamiltonian for describing
hadronic $B$ decays is given by
\be
{\cal H}_{eff} &=& \frac{4 G_F}{\sqrt{2}} V_{ts}^* V_{tb}
       \left[
           \sum_{i=1}^{2}
             \left( C_i(\mu) O_i(\mu) + C'_i(\mu) O'_i(\mu) \right)
           -\sum_{i=3}^{10}
             \left( C_i(\mu) O_i(\mu) + C'_i(\mu) O'_i(\mu) \right)
       \right],
\nonumber \\
&& ~~~ + {\rm H.c.}~,
\ee
including effects of the anomalous right-handed top quark
interactions.
The operator basis is defined following Ref. \cite{buras} by
\be
O_1 &=& (\bar{s}_\alpha c_\beta)_L (\bar{c}_\beta b_\alpha)_L,
\nonumber \\
O_2 &=& (\bar{s} c)_L (\bar{c} b)_L,
\nonumber \\
O_3 &=& (\bar{s} b)_L \sum_{q'=u,d,s,c,b} (\bar{q'} q')_L,
\nonumber \\
O_4 &=& (\bar{s}_\alpha b_\beta)_L \sum_{q'=u,d,s,c,b} (\bar{q'_\beta} q'_\alpha)_L,
\nonumber \\
O_5 &=& (\bar{s} b)_L \sum_{q'=u,d,s,c,b} (\bar{q'} q')_R,
\nonumber \\
O_6 &=& (\bar{s}_\alpha b_\beta)_L \sum_{q'=u,d,s,c,b} (\bar{q'_\beta} q'_\alpha)_R,
\nonumber \\
O_7 &=& \frac{3}{2}(\bar{s} b)_L \sum_{q'=u,d,s,c,b} e_{q'}(\bar{q'} q')_L,
\nonumber \\
O_8 &=& \frac{3}{2}(\bar{s}_\alpha b_\beta)_L 
                        \sum_{q'=u,d,s,c,b} e_{q'}(\bar{q'_\beta} q'_\alpha)_L,
\nonumber \\
O_9 &=& \frac{3}{2}(\bar{s} b)_L \sum_{q'=u,d,s,c,b} e_{q'}(\bar{q'} q')_R,
\nonumber \\
O_{10} &=& \frac{3}{2}(\bar{s}_\alpha b_\beta)_L 
                        \sum_{q'=u,d,s,c,b} e_{q'}(\bar{q'_\beta} q'_\alpha)_R,
\nonumber \\
O_{11} &=& \frac{g_s}{16 \pi^2} m_b 
           \bar{s}_\alpha P_R \sigma_{\mu \nu} T^a_{\alpha \beta} b_\beta G^{a~\mu \nu},
\nonumber \\
O_{12} &=& \frac{e}{16 \pi^2} m_b 
           \bar{s} P_R \sigma_{\mu \nu} b F^{\mu \nu},
\ee
where $(\bar{q} b)_{L/R} = (\bar{q} \gamma_\mu P_{L/R} b)$.
The operators $O'_i$ are the chiral conjugates
of the $O_i$ operators.

Matching the effective Hamiltonian and our model lagrangian of Eq. (2)
at $\mu = m_W$ scale,
we have the Wilson coefficients $C_i(\mu=m_W)$ and $C'_i(\mu=m_W)$
in the SM:
\be
C_1(m_W) &=& \frac{11}{2} \frac{\alpha_s(m_W)}{4 \pi},
\nonumber \\
C_2(m_W) &=& 1 - \frac{11}{6} \frac{\alpha_s(m_W)}{4 \pi} 
               - \frac{35}{18} \frac{\alpha}{4 \pi},
\nonumber \\
C_3(m_W) &=&  - \frac{\alpha_s(m_W)}{24 \pi} E_0(x_t)
              + \frac{\alpha}{6 \pi} \frac{1}{\sin^2 \theta_W}
                [2~B_0(x_t) + C_0(x_t)],
\nonumber \\
C_4(m_W) &=&  \frac{\alpha_s(m_W)}{8 \pi} E_0(x_t),
\nonumber \\
C_5(m_W) &=&  - \frac{\alpha_s(m_W)}{24 \pi} E_0(x_t),
\nonumber \\
C_6(m_W) &=&  \frac{\alpha_s(m_W)}{8 \pi} E_0(x_t),
\nonumber \\
C_7(m_W) &=&  \frac{\alpha}{6 \pi} [4~C_0(x_t) + D_0(x_t)],
\nonumber \\
C_9(m_W) &=&  \frac{\alpha}{6 \pi} [4~C_0(x_t) + D_0(x_t)
              + \frac{1}{\sin^2 \theta_W} (10~B_0(x_t) -4~ C_0(x_t))],
\nonumber \\
C_8(m_W) &=&  C_{10}(m_W) = 0,
\nonumber \\
C_{11}(m_W) &=& G(x_t),
\nonumber \\
C_{12}(m_W) &=& F(x_t),
\nonumber \\
C'_i(m_W) &=& 0, ~~~~~~~~~~~i=1,\cdot \cdot \cdot,12,
\ee
where $B_0(x), C_0(x), D_0(x), E_0(x), F(x)$, and $G(x)$ are the well-known
Inami-Lim loop functions of which explicit forms are given 
in Refs. \cite{buras,inami}.
Turning on the right-handed $\bar{t} b W$
and $\bar{t} s W$ couplings,
we have the modification of loop functions in the Wilson coefficients $C_i$,
\be
D_0(x_t) &\to& D_0(x_t) + \xi_b \frac{m_b}{m_t} D_R(x_t),
\nonumber \\
E_0(x_t) &\to& E_0(x_t) + \xi_b \frac{m_b}{m_t} E_R(x_t),
\nonumber \\
F(x_t) &\to& F(x_t) + \xi_b \frac{m_t}{m_b} F_R(x_t),
\nonumber \\
G(x_t) &\to& G(x_t) + \xi_b \frac{m_t}{m_b} G_R(x_t),
\ee
and also have the new Wilson coefficients $C'_i$:
\be
C'_3(m_W) &=&  - \frac{\alpha_s(m_W)}{24 \pi} \xi_s \frac{m_b}{m_t} E_R(x_t),
\nonumber \\
C'_4(m_W) &=&  \frac{\alpha_s(m_W)}{8 \pi} \xi_s \frac{m_b}{m_t} E_R(x_t),
\nonumber \\
C'_5(m_W) &=&  - \frac{\alpha_s(m_W)}{24 \pi} \xi_s \frac{m_b}{m_t} E_R(x_t),
\nonumber \\
C'_6(m_W) &=&  \frac{\alpha_s(m_W)}{8 \pi} \xi_s \frac{m_b}{m_t} E_R(x_t),
\nonumber \\
C'_7(m_W) &=&  \frac{\alpha}{6 \pi} \xi_s \frac{m_b}{m_t} D_R(x_t),
\nonumber \\
C'_9(m_W) &=&  \frac{\alpha}{6 \pi} \xi_s \frac{m_b}{m_t} D_R(x_t),
\nonumber \\
C'_{11}(m_W) &=& \xi_s \frac{m_t}{m_b} G_R(x_t),
\nonumber \\
C'_{12}(m_W) &=& \xi_s \frac{m_t}{m_b} F_R(x_t),
\nonumber \\
C'_i(m_W) &=&   0,~~~~( i = 1, 2, 8, 10 ),
\ee
where the new loop functions are given by
\be
D_R(x) &=& \frac{x(59-38x+25x^2+2x^3)}{36(x-1)^4}
        + \frac{2(x+1)}{3(x-1)^5} \ln x + \frac{x^2}{2(x-1)^4} \ln x,
\nonumber \\
E_R(x) &=& \frac{x(-116+165x-114x^2+29x^3)}{18(x-1)^4}
            + \frac{2+3x+x^2}{3(x-1)^5} \ln x,
\nonumber \\
F_R(x) &=& \frac{-20+31x-5x^2}{12(x-1)^2}
                 + \frac{x (2-3x)}{2(x-1)^3} \ln x,
\nonumber \\
G_R(x) &=& -\frac{4+x+x^2}{4(x-1)^2}
                 + \frac{3x}{2(x-1)^3} \ln x,
\ee
where our new loop functions $F_R(x)$, $G_R(x)$ and $D_R(x)$
agree with those in Ref. \cite{kylee,cho} and
$E_R(x)$ is the first calculation.
Note that the ${\cal O}(\xi)$ terms of $Z$--penguin diagram are
suppressed by the heavy mass of $Z$--boson as $m_b^2/m_Z^2$,
or $q^2/m_Z^2$ and we neglect them here.
For the box diagram, if we include only one anomalous couping,
the chirality structures of two currents are different
and one current is proportional to the fermion momentum and 
the other current proportional to the fermion mass.
It indicates that the integrand is always an odd function
and the loop integral vanishes. 
Therefore the ${\cal O}(\xi)$ terms of the box diagram
do not exist and the leading contribution is of $\xi^2$ order.
Hence, we also ignore the box contribution.
As a consequence, the contribution of order ${\cal O}(\xi)$
comes only through the $\gamma$-penguin and gluon penguin diagrams.
Actually the contributions of $O^{(\prime)}_{12}$ operator
to hadronic decays are very small and 
we neglect it in the numerical analysis.

The renormalization group (RG) evolution
of the Wilson coefficients ${\bf C} = (C_i,C'_i)^\dagger$ given by
\be
\mu \frac{d}{d \mu} {\bf C}(M_W)
  = -\frac{g_s^2}{16 \pi^2} \gamma^T {\bf C}(M_W),
\nonumber
\ee
is governed by a $24 \times 24$ anomalous dimension matrix $\gamma$.
Since the strong interaction preserves chirality,
the operators $O_i$ and $O'_i$ are evolved separately
without mixing between them.
Thus the $24 \times 24$ anomalous dimension matrix $\gamma$ is decomposed
into two identical $12 \times 12$ matrices $\gamma_{0}$ given in the SM.
The $12 \times 12$ anomalous dimension matrix $\gamma_0$
can be found in Ref. \cite{buras,ciuchini,buras0}.
The evolved Wilson coefficients $C^{(\prime)}_i(\mu)$
are expressed in terms of the initial conditions of Eq. (5) and (7),
$C(\mu) = U(\mu, M_W)C(M_W)$.
The explicit formula for the evolution matrix $U(\mu, M_W)$
can be found in Ref. \cite{ciuchini,buras0}. 
The matrix elements of operators also have one loop corrections.
We define the effective Wilson coefficients 
by absorbing the correction of the matrix elements 
in the Wilson coefficients
as given in Ref. \cite{he1,fleischer,alikramer}.
Then the Hamiltonian is expressed in terms of effective Wilson coefficients
and tree level matrix elements.

\section{$B \to X_{_s} \gamma$ Constraints}

Before the analysis on $\sin 2 \beta$, we consider the radiative 
$B \to X_s \gamma$ decay to constrain the model.
This channel has already been observed experimentally
and more precise measurement will be obtained from the
accumulation of data at $B$ factories.
It is well known that this process is an effective probe
of new physics since the dominant penguin diagram is
sensitive to the internal heavy particle property.
Especially, the right-handed couplings inside the loop 
of the operators $O_{11}$ and $O_{12}$ 
involve an enhancement factor $m_t/m_b$.
Thus the stringent limits on $\xi_b$ and $\xi_s$ are yielded
from the measurement of the $B \to X_s \gamma$ decay
\cite{kylee,larios}.
We present the updated constraints on anomalous couplings
from the branching ratio and the bound of CP violating asymmetry
in $B \to X_s \gamma$ decay.

The weighted average of the branching ratio is given by
\be
Br(B \to X_s \gamma) = (3.23 \pm 0.41 ) \times 10^{-4},
\ee
from the measurements of Belle \cite{belle2},
CLEO \cite{cleo1} and ALEPH \cite{aleph} groups.
The CP violating asymmetry 
in the $B \to X_s \gamma$ decays defined as
\be
A_{CP}(B \to X_s \gamma) 
= \frac{\Gamma(\bar{B} \to X_s \gamma) - \Gamma(B \to X_{\bar{s}} \gamma)}
       {\Gamma(\bar{B} \to X_s \gamma) + \Gamma(B \to X_{\bar{s}} \gamma)}
\nonumber
\ee
is very small in the SM 
because of the unitarity of the CKM matrix.
The direct CP asymmetry $A_{CP}$ is measured by CLEO \cite{cleo2}
\be
A_{CP}(B \to X_s \gamma) = (-0.079 \pm 0.108 \pm 0.022)(1.0 \pm 0.030),
\ee
where the first error is statistical, the second is additive systematic
over the various $b \to s \gamma$ decay modes,
and the third is multiplicative systematic.
Note that the present measurement of  $A_{CP}$ is still consistent
with 0.
The complex anomalous $\bar{t} b W$ coupling 
contributes to the CP asymmetry
through the interference terms of $O_{11}$ and $O_{12}$ operators such as 
$\delta A_{CP} \sim a_{1} Im C_2 C_{12}^* + a_{2} Im C_{11} C_{12}^* 
+ a_{3} Im C_2 C_{11}^*$, which provides an additional test on $\xi_b$
independent of the branching ratio.
On the contrary, the $\bar{t} s W$ coupling does not contribute to $A_{CP}$
at this level.

The explicit expressions of branching ratio and the CP asymmetry 
are presented in Refs. \cite{kn1} and \cite{kn2}
in terms of the evolved Wilson coefficients at $\mu=m_b$ scale.
With the measured values of Eqs. (9) and (10),
we obtain the constraints on $\xi_b$ at 2$\sigma$ C.L. as
\be
-0.002 < Re \xi_b + 22 | \xi_b |^2 < 0.0033,
\nonumber 
\ee
\be
-0.299 < \frac{\displaystyle 0.27~Im \xi_b}
         {\displaystyle 0.095 + 12.54 Re \xi_b + 414.23 |\xi_b|^2}  < 0.141,
\ee
and the allowed parameter set $(Re \xi_b,Im \xi_b)$ is depicted in Fig. 1.
Since $\xi_s$ is irrelevant for the CP asymmetry,
we can set the limit on $\xi_s$ to be
\be
| \xi_s | < 0.012,
\ee
from the branching ratio alone \cite{kylee}.

\section{Hadronic decays}

\subsection{$B \to J/\psi K$} 

The $B \to J/\psi K$ decays are dominated by the tree-level
$b \to c \bar{c} s$ decay amplitude and a single weak phase in the SM.
The subleading penguin contribution depends on the CKM factor
$V_{tb} V_{ts}^*$ which gives the same phase as the factor
$V_{cb} V_{cs}^*$ of the tree diagram
and the weak phase structure is not affected.
On that account, this mode is thought to be a golden mode
to extract the weak phase $\beta$.
The CP asymmetries in $B \to J/\psi K$ decays
given in Eq. (2), 
$\sin 2\beta_{\rm eff} = 0.734 \pm 0.054$,
agrees well with the SM prediction.
The subscript ``eff'' denotes the ``observed'' $\sin 2\beta$.
In terms of the Wilson coefficients,
the decay amplitude for $B \to J/\psi K_S$ decay is dominated by $C_2$ 
which involves no $\xi_{b,s}$ effects.
The subdominant amplitude involving $\xi_{b,s}$
is suppressed by loop suppression and/or a CKM factor
as well as the $\xi_{b,s}$ itself,
of which suppression factor is estimated of order $< 10^{-4}$.
Thus new physics effect on decay amplitude is ignored
to a very good approximation.
Considering the $B - \bar{B}$ mixing with right-handed coupling,
${\cal O}(\xi_b)$ contributions vanish in the box diagram calculation 
by the chirality relation and
the leading new physics contribution to the off-diagonal matrix 
element is of ${\cal O}(\xi_b^2)$, 
\be
M_{12} = M_{12}^{SM} \left( 1 + \xi_b^2 \frac{S_R(x_t)}{S_0(x_t)}
        \frac{(\bar{b}P_Ld)(\bar{b}P_Ld)}
           {(\bar{b}\gamma_\mu P_Ld)(\bar{b}\gamma^\mu P_Ld)} \right),
\ee
where the new loop function $S_R(x)$ is given by
\be
S_R(x)=\frac{x(x^2-2x+6)}{(1-x)^2} + \frac{x(x+2)(x^2-x+2)}{(1-x)^3}\ln x,
\ee
and the SM loop function $S_0(x)$ can be found in Ref. \cite{buras}.
It leads to the ${\cal O}(\xi_b^2)$ shift of the weak phase
\be
\sin 2 \beta_{\rm eff} = \sin 2 \beta + 4.3 |\xi_b|^2 \sin 2 \varphi,
\ee
where $\xi_b = |\xi_b| e^{i \varphi}$ and
\be
\frac{\langle B^0 |(\bar{b}P_Ld)(\bar{b}P_Ld) | \bar{B}^0 \rangle}
    {\langle B^0 |(\bar{b}\gamma_\mu P_Ld)
                     (\bar{b}\gamma^\mu P_Ld) | \bar{B}^0 \rangle} 
      \approx \frac{3}{4} \left( \frac{m_B}{m_b} \right)^2.
\nonumber
\ee
With the allowed parameter set of Fig. 1,
the second term of Eq. (16) is at most of order $10^{-3}$
so we can neglect it for the discussion of $\sin 2\beta$.
On the other hand, the anomalous $\bar{t}sW$ coupling is irrelevant for
the $B - \bar{B}$ mixing and invokes no effects on $\sin 2\beta_{\rm eff}$
in $ B \to J/\psi K$ decays.
As a consequence, the observed weak phase $\sin 2 \beta$ 
in $B \to J/\psi K$ decays is hardly affected by
the right-handed top couplings.

\subsection{$B \to \phi K_S$} 

The average of the CP asymmetry in $B \to \phi K_S$ decay 
measured by BaBar \cite{babar1} and Belle \cite{belle1} groups
is given by
\be
&A_{CP}^{\phi K} = -0.56 \pm 0.43,&
\nonumber \\ 
&\sin 2 \beta_{\rm eff}^{\phi K} = -0.39 \pm 0.41,&
\ee
where $A_{CP}^{\phi K}$ is the CP violating asymmetry defined as
$A_{CP}^{\phi K} \equiv 
[ \Gamma(B \to \phi K) - \Gamma(\bar{B} \to \phi \bar{K}) ]
/[ \Gamma(B \to \phi K) + \Gamma(\bar{B} \to \phi \bar{K}) ]
$
and $\sin 2 \beta_{\rm eff}^{\phi K}$ the observed weak phase
extracted from $B \to \phi K_S$ decay.

The $b \to s \bar{s} s$ transition responsible for the
$B \to \phi K$ decays arises at one loop level in the SM.
It is known that the gluon penguin diagram plays a central role
in this decay channel through the chromo-magnetic (dipole penguin) 
operator $O_{11}$ as well as the four quark operator.
As in the case of $B \to X_s \gamma$ decay,
if the right-handed couplings are switched on,
the enhancement factor $m_t/m_b$ involved by the penguin loop
makes the new effect of $O^{(\prime)}_{11}$ operator
lead to significant contributions in $B \to \phi K$ decays.
It has been discussed that the anomalous right-handed
$\bar{t} b W$ coupling can yield a deviation of the CP asymmetry
in $B \to \phi K_S$ process from the SM prediction 
by Abd El-Hady and Valencia \cite{elhady}.
Here we present the detailed analysis on the CP violation
in $B \to \phi K_S$ decays
including both of $\bar{t} b W$ and $\bar{t} s W$ couplings
and compare $A_{CP}^{\phi K}$ and
$\sin 2 \beta_{\rm eff}^{\phi K}$ with the experiment.
On the other hand, the electroweak penguin operators
also give sizable contribution to this decay mode, up to 20\% \cite{he1}.
Therefore we include all operators in the effective Hamiltonian
to evaluate the $B \to \phi K_S$ decay rate
except for $O_{12}$ since its contribution is extremely small.

With the definition of the form factors and decay constants
\be
\langle P(p') | V_\mu | B(p) \rangle &=&
\left[ (p+p')_\mu - \frac{m_B^2-m_P^2}{q^2} q_\mu \right] F_1^P(q^2)
+ \frac{m_B^2-m_P^2}{q^2} q_\mu F_0^P(q^2),
\nonumber \\
\langle 0 | A_\mu | P(p) \rangle &=& i f_P p_\mu,
\nonumber \\
\langle 0 | V_\mu | V(p) \rangle &=& f_V m_V \epsilon_\mu,
\ee
we write the decay amplitude for $B \to \phi K$ decays as
\be
{\cal A}(B^0 \to \phi K^0) = -\frac{G_F}{\sqrt{2}} V_{tb}^* V_{ts}
\left[ a_3 + a_4 + a_5 - \frac{1}{2} (a_7 + a_9 + a_{10}) \right]
2 f_\phi m_\phi (\epsilon^* \cdot p_B) F_1^K + A_{11}^{\phi K},
\ee
where $a_{2i-1} = C_{2i-1} +C_{2i}/N_c$, $a_{2i} = C_{2i} +C_{2i-1}/N_c$.
Contribution of the chromo-magnetic operator $A_{11}^{\phi K}$ is given by 
\cite{kagan,he2}
\be
A_{11}^{\phi K} &\equiv &
\langle \phi K^0 | ({\cal H}_{11} + {\cal H'}_{11}) | B^0 \rangle,
\nonumber \\
 &=& \frac{G_F}{\sqrt{2}} \frac{\alpha_s(q^2)}{4 \pi q^2} V_{tb}^* V_{ts}
      m_b(\mu) \frac{N_c^2-1}{N_c^2} f_\phi m_\phi (\epsilon^* \cdot p_B)
      (C_{11} + C'_{11})
     ( F_1^K X + F_0^K Y),
\ee
with
\be 
X &=& 4 m_b + 5 m_s + 3 m_s \left( \frac{m_B^2-m_K^2}{m_\phi^2} \right) 
      - \left( \frac{3 m_B^2 -3 m_K^2 + m_\phi^2}{8 m_b} \right) 
        \left( 1 + \frac{m_B^2-m_K^2}{m_\phi^2} \right),
\nonumber \\
Y &=& \frac{3}{2} \left( \frac{m_B^2-m_K^2}{m_b - m_s} \right)
      + \left( \frac{m_B^2-m_K^2}{m_\phi^2} \right)
        \left( \frac{3 m_B^2 -3 m_K^2 + m_\phi^2}{8 m_b} -3 m_s \right),
\ee
and $q^2 = m_B^2/2 - m_K^2/4 + m_\phi^2/2$.
The $B$ to $K$ form factor, $F_{0,1}$ is the principal source of 
hadronic uncertainty for this process.
The early calculation was performed in the framework of a quark model
\cite{wsb}.
We can set $F_0 = F_1 $ close to the point $q^2 =0$ \cite{ali}
and assume the simple pole-dominance.
Here we take the value of $ F_{0,1}(0) = 0.26 - 0.37$ 
from the QCD sum rule results \cite{belyaev}. 
Note that new effects on four-quark operators
are doubly suppressed by both $m_b/m_t$ and $\xi_{b,s}$,
while the effects on dipole operator involve an
enhancement factor $m_t/m_b$ compensating the $\xi_{b,s}$
suppression.
Thus the new contribution dominantly comes through $A_{11}^{\phi K}$.
We also notify that $C^{(\prime)}_i$ in the Eqs. (19) and (20)
are the effective Wilson coefficients absorbing the 1-loop correction
to the hadronic matrix elements and  they involve the strong phases.

Since the four-quark operator contribution in the first term in Eq. (19)
involves a strong phase, the new phase of $A_{11}^{\phi K}$
leads to a deviation of $| \bar{A} / A |$ from unity
and we have the rate asymmetry implying the direct CP violation.
In terms of the parameter $\lambda$ defined by
\be
\lambda = \sqrt{ \frac{M_{12}^*}{M_{12}} } \frac{\bar{A}}{A},
\ee
where $A={\cal A}(B^0 \to \phi K^0)$ and
$\bar{A}={\cal A}(\bar{B}^0 \to \phi \bar{K}^0)$,
we write the full expression of the time-dependent CP asymmetry as
\be
a_{\phi K}(t) &\equiv& 
\frac{ \Gamma(B^0_{\rm phys}(t) \to \phi K^0) 
             - \Gamma(\bar{B}^0_{\rm phys}(t) \to \phi \bar{K}^0) }
{ \Gamma(B^0_{\rm phys}(t) \to \phi K^0)
             + \Gamma(\bar{B}^0_{\rm phys}(t) \to \phi \bar{K}^0) },
\nonumber \\
&=& C_{\phi K} \cos \Delta m_B t + S_{\phi K} \sin \Delta m_B t,
\ee
where the coefficients are
\be
C_{\phi K} &= & \frac{1-|\lambda|^2}{1+|\lambda|^2}
             \equiv - A_{CP}^{\phi K},
\nonumber \\
S_{\phi K} &= & -\frac{2 Im \lambda}{1+|\lambda|^2}
             \equiv \sin 2 \beta_{\rm eff}^{\phi K}.
\ee
Note that the hadronic uncertainty is cancelled in $\lambda$
and the CP violating observables $A_{CP}^{\phi K}$ and $\sin 2 \beta_{\rm eff}^{\phi K}$
are free from the hadronic uncertainty.
We can express the parameter $\lambda$ by
\be
\lambda = \lambda^{\rm SM}
          \left( \frac{1+21.84~e^{-i \delta}~\xi_q}   
                      {1+21.84~e^{-i \delta}~\xi_q^*}   
          \right)
\approx \lambda^{\rm SM} (1 + i~43.7~|\xi_q|~e^{-i \delta} \sin \varphi_q )
\ee
where $\varphi_q$ is the phase of $\xi_q$,
and $\delta$ the strong phase introduced by the one loop 
correction to the matrix elements.
For $\lambda^{\rm SM} \equiv e^{i \beta_{\rm SM}}$, we will use
the measured value given in Eq. (2).
With this expression, we can write the CP asymmetries as
\be
A_{CP}^{\phi K} &=& 23.3~|\xi_q|~\sin \varphi_q,
\nonumber \\
\sin 2 \beta_{\rm eff}^{\phi K} &=& \sin 2 \beta + 52.2 ~|\xi_q|~\sin \varphi_q,
\ee
where $\delta = 2.58$ in our calculation.
Note that the second expression of Eq. (25) is no more valid
for the maximal value of $|\xi_b| \sim 0.04$ 
and so are the above expressions of $A_{CP}^{\phi K}$ and 
$\sin 2 \beta_{\rm eff}^{\phi K}$. 

With the allowed parameter set given in Fig. 1 and measured $\sin 2 \beta$
in $B \to J/\psi K$ decay given in Eq. (13), we have
the rate asymmetry $A_{CP}^{\phi K}$ and the effective weak phase
$\sin 2 \beta_{\rm eff}^{\phi K}$ :
\be
&-0.34 < A_{CP}^{\phi K} < 0.22,&
\nonumber \\
&-0.10 < \sin 2 \beta_{\rm eff}^{\phi K} < 0.96.&
\ee
As shown in the previous section, the effect of the right-handed top
couplings on $B - \bar{B}$ mixing sector is safely neglected
for the evaluation of $\sin 2 \beta_{\rm eff}$.
In Fig. 2, the correlation of $\sin 2 \beta_{\rm eff}^{\phi K}$
and $A_{CP}^{\phi K}$ is shown.
We find that a large rate asymmetry ($-20 \sim -30$ \%) should exist
for the observed $\sin 2 \beta$ to be negative.
Even if the future experiment ascertain that
the $\sin 2 \beta_{\rm eff}^{\phi K}$ is consistent with the SM prediction,
it is still possible that there exists a sizable direct CP
violation $A_{CP}^{\phi K} \sim 10$ \%.
With the right-handed $\bar{t} s W$ coupling $| \xi_s | < 0.012$,
we have
\be
& -0.28 < A_{CP}^{\phi K} < 0.28,&
\nonumber \\
& 0.23 < \sin 2 \beta_{\rm eff}^{\phi K} < 0.94,&
\ee
and their correlation is shown in Fig. 3.
We also find that the large CP asymmetry ($\sim 10$ \%) is possible with $\xi_s$
even if $\sin 2 \beta_{\rm eff}$ agrees with the SM prediction.

We also calculate the branching ratio :
\be
Br(B \to \phi K) = \tau_B \frac{1}{16 \pi} 
                       \frac{\lambda(m_B^2,m_\phi^2,m_K^2)}{m_B^3}
                       | {\cal A} |^2,
\ee
where $\lambda(x,y,z) = (x^2+y^2+z^2 -2xy -2yz -2zx)^{1/2}$
and $\tau_B$ is the lifetime of $B$ meson.
Figure 4 and 5 show the relations of the branching ratio and CP violations
in the presence of the right-handed $\bar{t}bW$ and $\bar{t}sW$ couplings.
The present measurements of the branching ratio for $B^0 \to \phi K^0$
decay 
read
\be
{\rm Br}(B^0 \to \phi K^0) 
&=& (5.4^{+3.7}_{-2.7}\pm0.7)\times 10^{-6} 
                 < 12.3\times 10^{-6}~~~~~~{\rm CLEO},
\nonumber \\
&=& (8.1^{+3.1}_{-2.5}\pm0.8)\times 10^{-6} 
                 ~~~~~~~~~~~~~~~~~~~~~~~~~{\rm BaBar} ,
\nonumber \\
&=& (8.7^{+3.8}_{-3.0}\pm1.5)\times 10^{-6} 
                 ~~~~~~~~~~~~~~~~~~~~~~~~~{\rm Belle} ,
\ee
by the CLEO \cite{cleobr}, BaBar \cite{babarbr} and 
Belle \cite{bellebr} groups. 
Since the CLEO result is just an intermediate fitted value
and the Belle result is a preliminary one,
we just show the BaBar result in the Fig. 4 and 5.
In our evaluation, the SM value is $(1.9-4.0)\times10^{-6}$
close to the prediction of Ref. \cite{chengbr}.
We can find that the negative $\sin 2 \beta_{\rm eff}$ 
consistent with the BaBar cross section is possible 
with anomalous $\bar{t}bW$ coupling 
but we do not expect such a solution with anomalous $\bar{t}sW$ coupling.

We also show the correlation of the CP asymmetries
between $ B \to X_s \gamma$ and $B \to \phi K$ decays 
in Fig. 6.
For the negative $\sin 2 \beta_{\rm eff}$ in $B \to \phi K$ decay, 
$-(2-3)$ \% of $A_{CP}^\gamma$ is expected.

\section{Concluding Remarks} 
 
We have studied the effects of the complex right-handed top quark couplings 
on the CP violation in $B \to \phi K$ decays,
which originate in the general
SU(2)$_L \times$SU(2)$_R \times$U(1) model or 
the dynamical electroweak symmetry breaking model. 
Since the contribution of those couplings to the $B-\bar{B}$ mixing
is suppressed by the quadratic order of $\xi_q$,
the measurement of the $\sin 2 \beta$ in $B \to J/\psi K$ decays
is not affected by the right-handed couplings.
However, the gluonic dipole penguin operator, which
plays a important role in $b \to s \bar{s} s$ decay, 
gets a sizable contribution from the right-handed couplings 
due to an enhancement factor $m_t/m_b$ 
and the observed $\sin 2 \beta$ in $B \to \phi K$ decays can be shifted.
Even the negative $\sin 2 \beta$ is possible 
with the anomalous $\bar{t}bW$ coupling,
as is consistent with the recent measurements.
Note that the $\sin 2 \beta$ with the anomalous $\bar{t}sW$ coupling
is also shifted but still positive 
due to more strict bound on $|\xi_s|$ than that on $|\xi_b|$.
In conclusion, the right-handed top couplings are 
one of the good candidate of the present disagreement 
of the observed $\sin 2 \beta$
between $B \to J/\psi K$ and $B \to \phi K$ decays
if it exists.

Furthermore, since the complex phase of right-handed couplings
is a new source of CP violation, 
the rate asymmetry indicating a direct CP violation
also exists in $B \to \phi K$ decays. 
This CP asymmetry may be large, up to $-30$ \% and
another strong indication of the right-handed top couplings.

\acknowledgments 
 
\noindent 
We thank S. Oh for valuable comments. 
KYL thanks S.W. Baek for helpful discussions.
The research of J.P. Lee was supported by the BK21 Program.

\def\PRD #1 #2 #3 {Phys. Rev. D {\bf#1},\ #2 (#3)} 
\def\PRL #1 #2 #3 {Phys. Rev. Lett. {\bf#1},\ #2 (#3)} 
\def\PLB #1 #2 #3 {Phys. Lett. B {\bf#1},\ #2 (#3)} 
\def\NPB #1 #2 #3 {Nucl. Phys. {\bf B#1},\ #2 (#3)} 
\def\ZPC #1 #2 #3 {Z. Phys. C {\bf#1},\ #2 (#3)} 
\def\EPJ #1 #2 #3 {Euro. Phys. J. C {\bf#1},\ #2 (#3)} 
\def\IJMP #1 #2 #3 {Int. J. Mod. Phys. A {\bf#1},\ #2 (#3)} 
\def\MPL #1 #2 #3 {Mod. Phys. Lett. A {\bf#1},\ #2 (#3)} 
\def\PTP #1 #2 #3 {Prog. Theor. Phys. {\bf#1},\ #2 (#3)} 
\def\PR #1 #2 #3 {Phys. Rep. {\bf#1},\ #2 (#3)} 
\def\RMP #1 #2 #3 {Rev. Mod. Phys. {\bf#1},\ #2 (#3)} 
\def\PRold #1 #2 #3 {Phys. Rev. {\bf#1},\ #2 (#3)} 
\def\IBID #1 #2 #3 {{\it ibid.} {\bf#1},\ #2 (#3)}

\smallskip 
\smallskip 
\smallskip 
 
\begin{center} 
\begin{figure}[htb] 
\hbox to\textwidth{\hss\epsfig{file=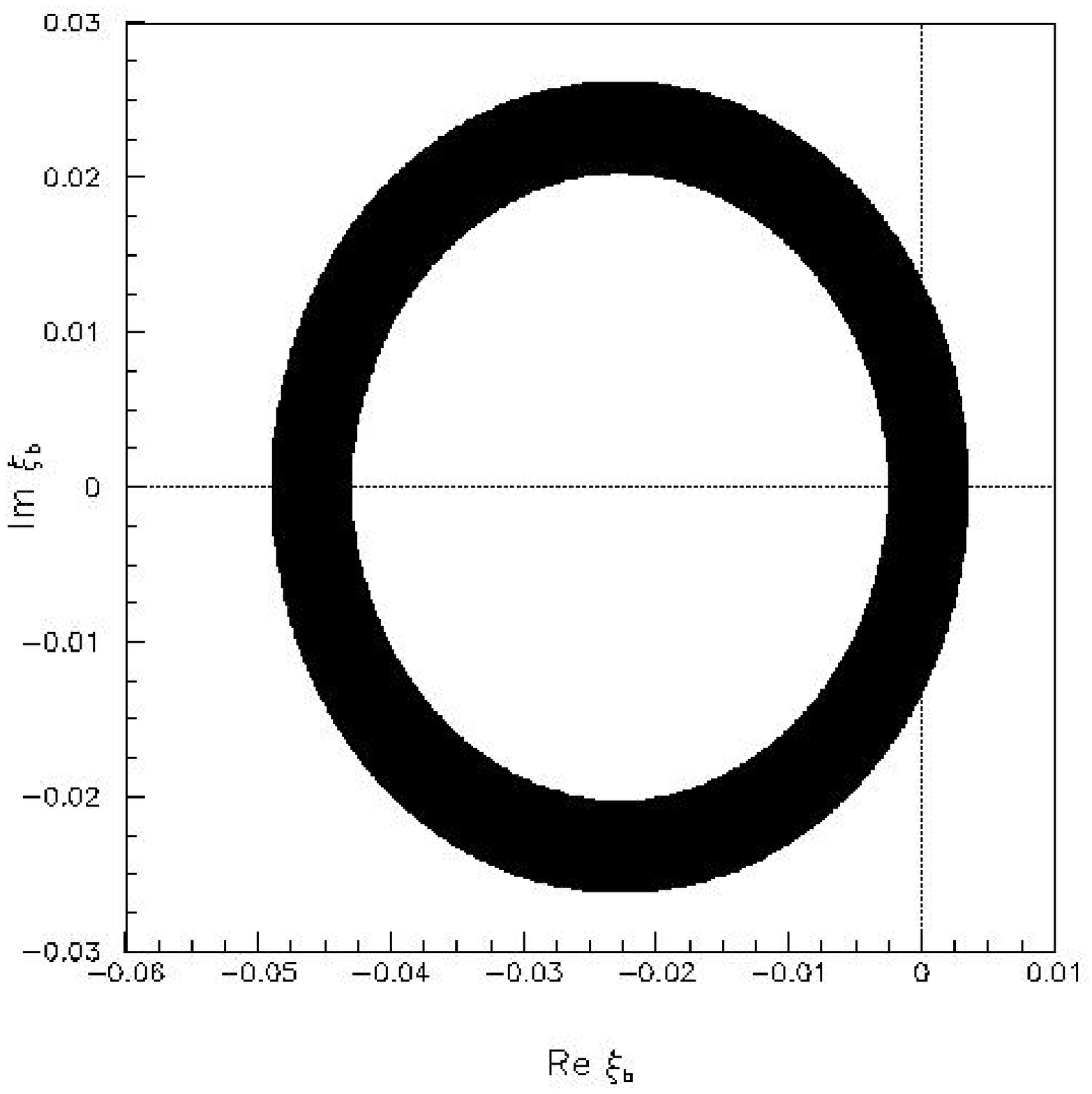}\hss} 
\label{fig1} 
\caption{ Allowed parameter set $(Re \xi_b,Im \xi_b)$
under the constraints by the branching ratio and CP asymmetry
in $B \to X_s \gamma$ decay.
} 
\end{figure} 
\end{center} 
 
\newpage 
 
\begin{center} 
\begin{figure}[htb] 
\hbox to\textwidth{\hss\epsfig{file=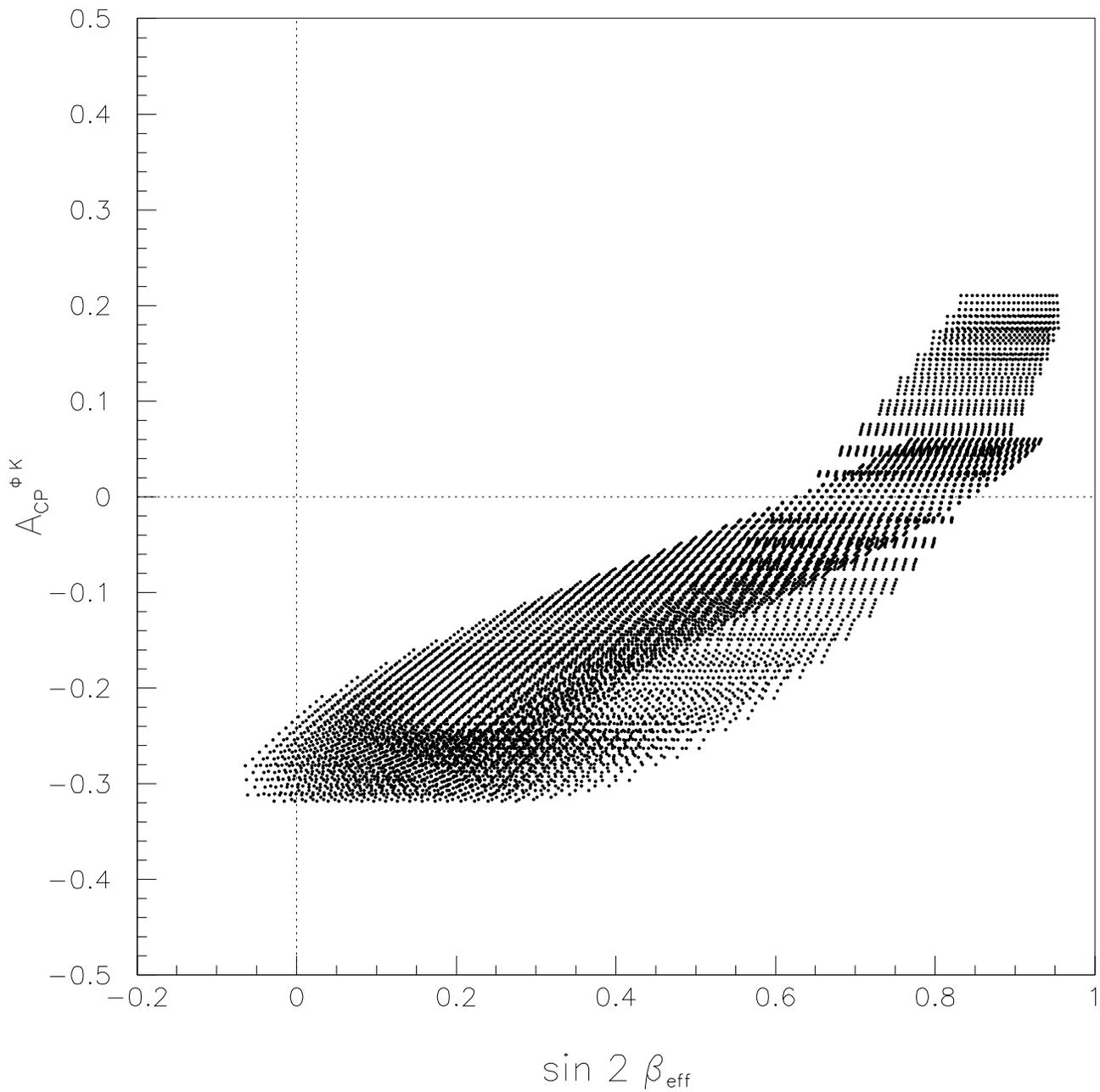}\hss} 
\caption{Correlation of $\sin 2 \beta_{\rm eff}$ and $A_{CP}^{\phi K}$
with varying $\xi_b$ shown in Fig. 1.
} 
\label{fig2} 
\end{figure} 
\end{center} 
 
\newpage 
 
\begin{center} 
\begin{figure}[htb] 
\hbox to\textwidth{\hss\epsfig{file=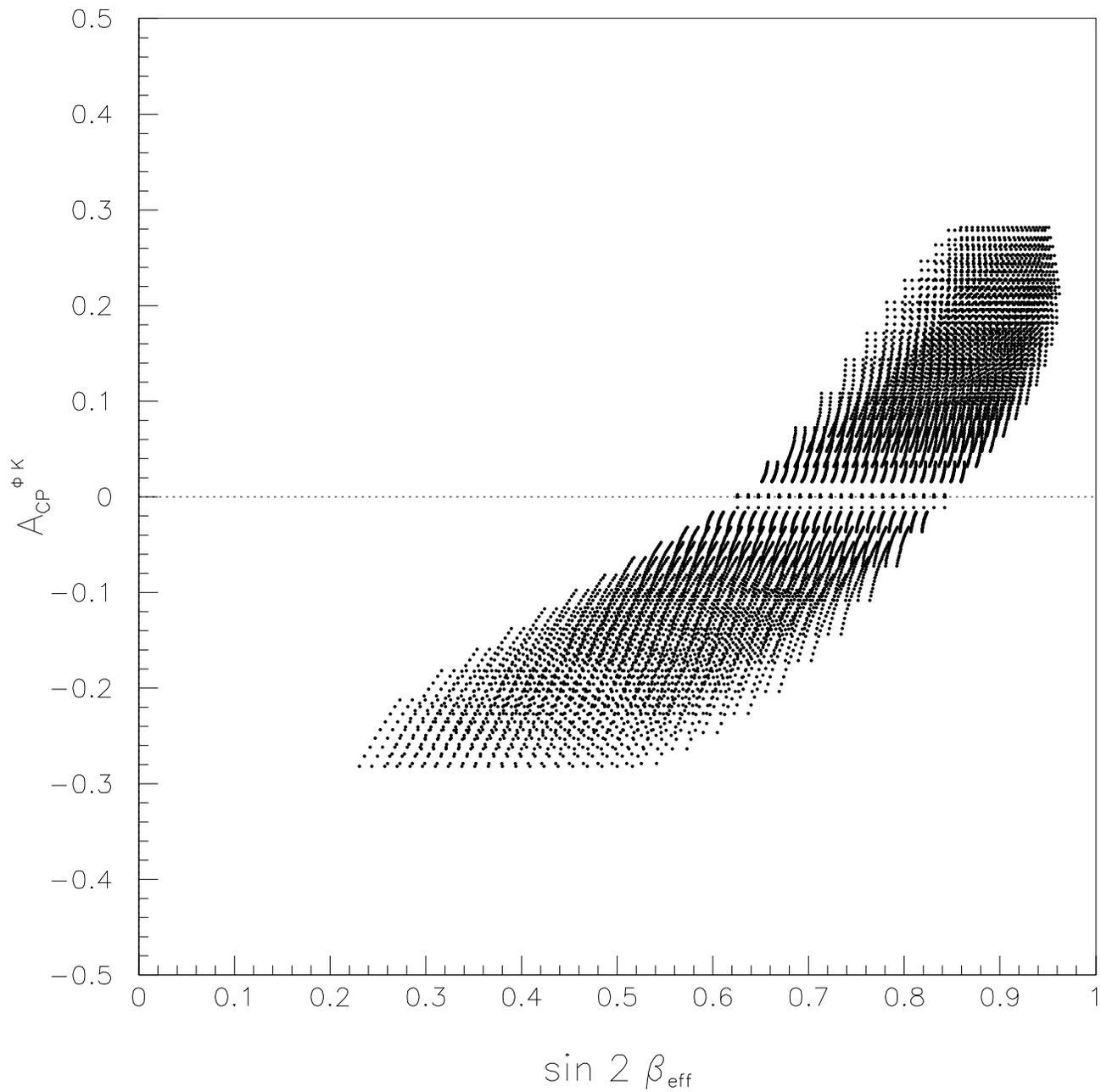}\hss} 
\caption{Correlation of $\sin 2 \beta_{\rm eff}$ and $A_{CP}^{\phi K}$
with varying $\xi_s$ under the constraint of Eq. (11).
}
\label{fig3} 
\end{figure} 
\end{center} 
 
\newpage 
 
\begin{center} 
\begin{figure}[htb] 
\hbox to\textwidth{\hss\epsfig{file=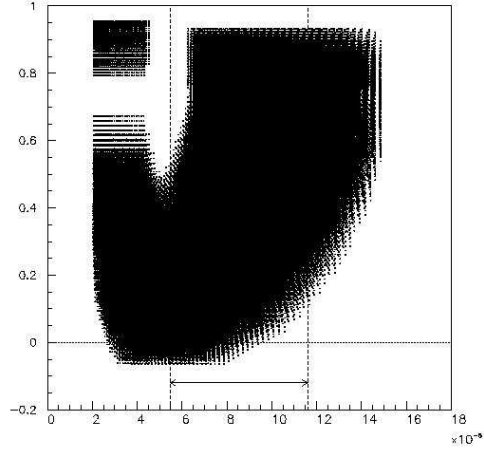,height=10cm}\hss} 
\label{fig4a} 
 
\hbox to\textwidth{\hss\epsfig{file=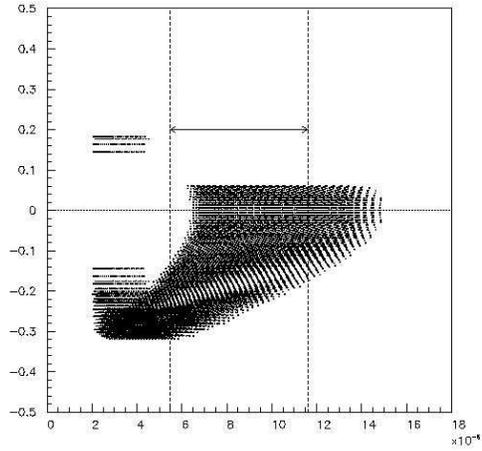,height=10cm}\hss} 
\vskip 0.5cm 
\caption{
(a) Correlation of the branching ratio of $B \to \phi K$ decay and
$\sin 2 \beta_{\rm eff}^{\phi K}$ with varying $\xi_b$ shown in Fig. 1. 
(b) Correlation of the branching ratio of $B \to \phi K$ decay and 
$A_{CP}^{\phi K}$ with varying $\xi_b$ shown in Fig. 1.
} 
\label{fig4b} 
\end{figure} 
\end{center} 
 
\newpage 
 
\begin{center} 
\begin{figure}[htb] 
\hbox to\textwidth{\hss\epsfig{file=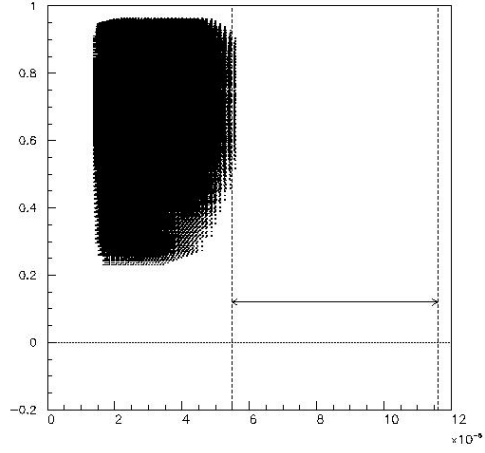,height=10cm}\hss} 
\label{fig5a} 
\hbox to\textwidth{\hss\epsfig{file=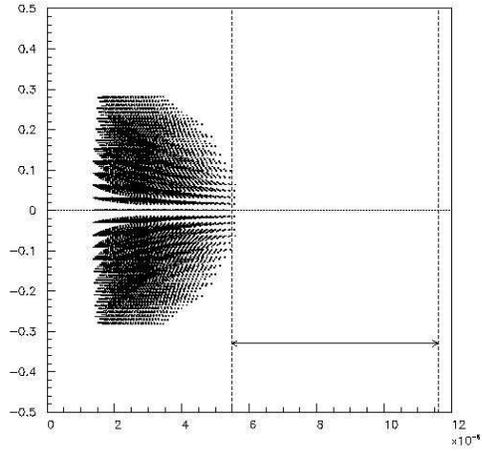,height=10cm}\hss} 
\vskip 0.5cm 
\caption{ 
(a) Correlation of the branching ratio of $B \to \phi K$ decay and
$\sin 2 \beta_{\rm eff}^{\phi K}$ 
with varying $\xi_s$ under the constraint of Eq. (11).
(b) Correlation of the branching ratio of $B \to \phi K$ decay and 
$A_{CP}^{\phi K} $
with varying $\xi_s$ under the constraint of Eq. (11).
} 
\label{fig5b} 
\end{figure} 
\end{center} 
 
\newpage 
 
\begin{center} 
\begin{figure}[htb] 
\hbox to\textwidth{\hss\epsfig{file=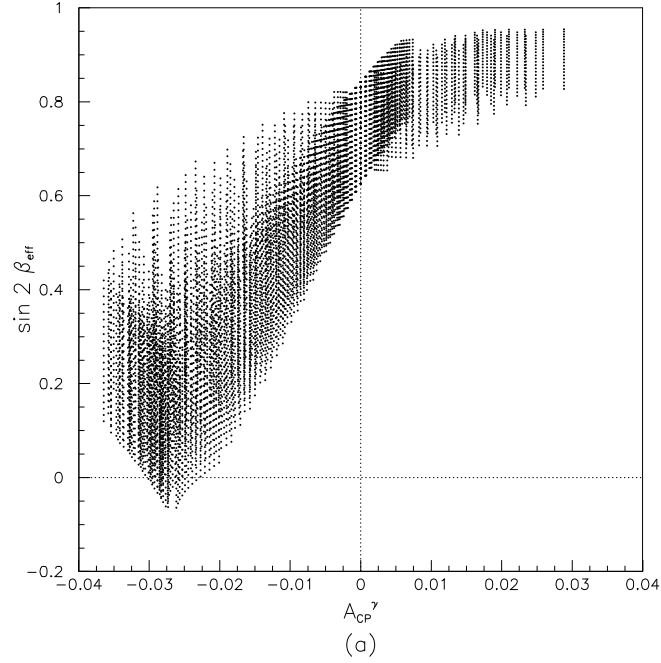,height=10cm}\hss} 
\label{fig6a} 
 
\hbox to\textwidth{\hss\epsfig{file=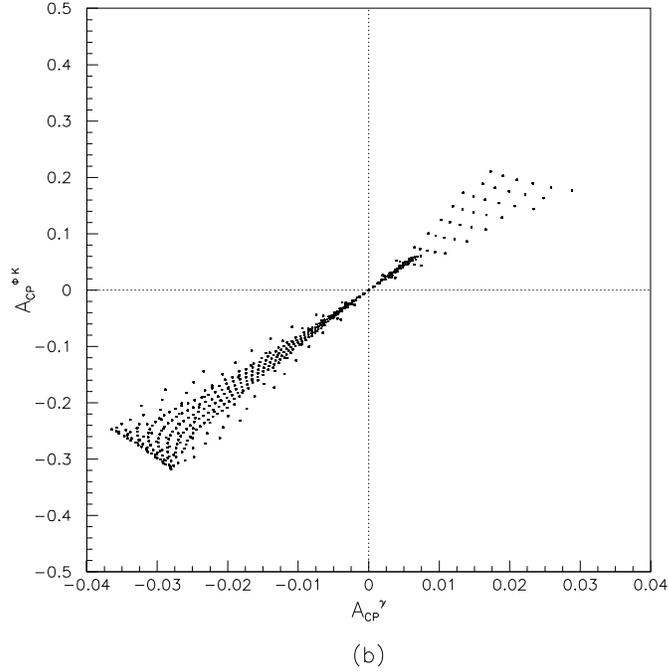,height=10cm}\hss} 
\vskip 0.5cm 
\caption{ 
(a) Correlation of the CP asymmetry of $B \to X_s \gamma$ decay and
$\sin 2 \beta_{\rm eff}^{\phi K}$ with varying $\xi_b$ shown in Fig. 1.
(b) Correlation of the CP asymmetry of $B \to X_s \gamma$ decay and
$A_{CP}^{\phi K} $with varying $\xi_b$ shown in Fig. 1.
} 
\label{fig6b} 
\end{figure} 
\end{center} 
 
\end{document}